\documentclass[twocolumn]{revtex4}

\usepackage{ragged2e}
\usepackage{amsmath}
\usepackage{graphicx}

\usepackage[usenames,dvipsnames]{color}

\definecolor{darkblue}{RGB}{0,0,196}

\begin{document}

\title{Thermodynamic and hydrodynamic characteristics of interacting system
formed in relativistic heavy ion collisions \vspace{0.5cm}}

\author{Xu-Hong~Zhang$^{1,}$\footnote{xhzhang618@163.com; zhang-xuhong@qq.com},
Hao-Ning~Wang$^{2,}$\footnote{wanghaoning517@139.com},
Fu-Hu~Liu$^{1,}$\footnote{Correspondence: fuhuliu@163.com;
fuhuliu@sxu.edu.cn},
Khusniddin~K.~Olimov$^{3,4,}$\footnote{Correspondence:
khkolimov@gmail.com; kh.olimov@uzsci.net}}

\affiliation{$^1$State Key Laboratory of Quantum Optics and
Quantum Optics Devices, Institute of Theoretical Physics, Shanxi
University, Taiyuan 030006, China
\\
$^2$College of Mechanical and Vehicle Engineering, Taiyuan
University of Technology, Taiyuan 030024, China
\\
$^3$Laboratory of High Energy Physics, Physical-Technical
Institute of Uzbekistan Academy of Sciences, Chingiz Aytmatov Str.
2b, Tashkent 100084, Uzbekistan
\\
$^4$Department of Natural Sciences, National University of Science
and Technology MISIS (NUST MISIS), Almalyk Branch, Almalyk 110105,
Uzbekistan}

\begin{abstract}

\vspace{0.5cm}

\noindent {\bf Abstract:} To study the energy-dependent
characteristics of thermodynamic and hydrodynamic parameters,
based on the framework of a multi-source thermal model, we analyze
the soft transverse momentum ($p_{T}$) spectra of the charged
particles ($\pi^{-}$, $\pi^{+}$, $K^{-}$, $K^{+}$, $\bar{p}$, and
$p$) produced in gold--gold (Au--Au) collisions at the
center-of-mass energies $\sqrt{s_{NN}}=7.7$, 11.5, 14.5, 19.6, 27,
39, 62.4, and 200~GeV from the STAR Collaboration and in
lead--lead (Pb--Pb) collisions at $\sqrt{s_{NN}}=2.76$ and
5.02~TeV from the ALICE Collaboration. In the rest framework of
emission source, the probability density function obeyed by meson
momenta satisfies the Bose-Einstein distribution, and that obeyed
by baryon momenta satisfies the Fermi-Dirac distribution. To
simulate the $p_{T}$ of the charged particles, the kinetic
freeze-out temperature $T$ and transverse expansion velocity
$\beta_{T}$ of emission source are introduced into the
relativistic ideal gas model. Our results, based on the Monte
Carlo method for numerical calculation, show a good agreement with
the experimental data. The excitation functions of thermodynamic
parameter $T$ and hydrodynamic parameter $\beta_{T}$ are then
obtained from the analyses, which shows an increase tendency from
7.7 GeV to 5.02 TeV in collisions with different centralities.
\\
\\
{\bf Keywords:} Multi-source thermal model, Transverse momentum
spectra, Bose-Einstein (Fermi-Dirac) distribution, Thermodynamic
and hydrodynamic parameters (characteristics)
\\
\\
{\bf PACS:} 12.40.Ee, 13.85.Hd, 24.10.Pa
\\
\\
\end{abstract}

\maketitle

\section{Introduction}

The strong interaction, which was used to describe the nuclear
force between nucleons (protons or neutrons) at the earliest stage
of the universe evolution, is the strongest one~\cite{1,2}, being
38~orders of magnitude greater than gravity. As it differs from
the atomic and molecular scale interactions, scientists usually
use quantum chromodynamics (QCD) to describe the strong
interactions between particles below the size of a
nucleus~\cite{3,4,5}. In recent years, the research on ``new state
of matter", formed in relativistic heavy ion collisions, has
attracted many scientists in the field of high energy physics in
the world~\cite{6}.

In extreme conditions, raising the temperature of the system, or
increasing its density, the QCD substance will produce two
particular phase transitions, one is the deconfinement phase
transition and the other is the chiral phase
transition~\cite{7,8}. The former corresponds to the disappearance
of quark confinement, meaning that quarks are free to move to
other regions of the nuclear matter, not just confined to the
movement inside the nucleon. The latter corresponds to the
restoration of eigensymmetry, that is, the kinetic mass becomes
zero and the quarks become as particles, close to zero-mass, while
at the same time, the existence of phase transitions indicates the
emergence of a new state of matter.

Observation of these two phase transitions in a high-temperature
dense region implies the transition of a substance to a state with
quark-gluons as the fundamental degree of freedom, and this new
state of matter is named quark-gluon plasma (QGP)~\cite{9,10,11}.
In QGP expansion, a lot of particles are produced, and finally
they are measured in experiments. One may study the formation and
change of the new state of matter from the dynamic evolution of
the final state particles. Such a matter is usually described by
QCD phase diagram~\cite{7,8}, and the thermodynamic properties of
the system are expressed by the temperature and the chemical
potential of the baryons.

When the system undergoes phase transition and reaches the
equilibrium state of chemical and kinetic freeze-out stage, the
thermodynamic properties of the final state particles can be
studied, which is of great significance to obtain the critical
point of phase transition and understand the characteristics of
QCD. In addition, under low temperature and high density
condition, QCD substance will form color superconducting
state~\cite{12,13,14} through phase transition. The above two new
states of matter do not exist stably, and the heavy ion colliders
can be used to control the system state, which provides a powerful
tool for researchers to study the QCD phase transition.

As an open question, the energy of the critical point is worth
studying by various ways~\cite{14a,14b,14c,14d,14e}. One of them
is to study the energy-dependent relations or the excitation
functions of different parameters which include, but are not
limited to, the thermodynamic parameter $T$ and the hydrodynamic
parameter $\beta_T$, Generally, an inflection point or sudden
change in the excitation functions implies the existence of
critical point or a change of interaction mechanism. Experiments
in relativistic heavy ion collisions~\cite{15,16} have provided a
new chance for ones to explore new matter and phenomena under
extreme conditions. Meanwhile, one may test different theories or
models and explain the new effects~\cite{17,18,19}.

The whole process of relativistic heavy ion collisions can be
divided into three stages: pre-equilibrium dynamics, viscous fluid
dynamics, and free flow. In the collisions, a large number of
particles are generated and escaped. At the last stage of free
flow (i.e. the stage of kinetic freeze-out), one is curious
whether the widely used relativistic ideal gas
model~\cite{35,36,37} can be applied. If yes, escaped fermions
should be subject to Fermi-Dirac statistics and bosons follow the
Bose-Einstein statistics. This can be done if one assumes that the
generated particles come from the equilibrium stationary source,
though the evolution process is represented by a perfect liquid.

However, because the process of relativistic heavy ion collisions
is very complex and the number of produced particles is very
large. It is possible that an equilibrium stationary source is not
enough to describe more characteristics. One naturally thinks of a
multi-source scenario, the multi-source thermal
model~\cite{33,34}, which assumes multiple equilibrium stationary
sources, and there are interactions among these sources. Due to
the extreme squeeze between projectile and target nuclei, a
transverse expansion of the interacting system (emission source)
or a transverse flow of the final state particles will appear,
which affects particle momentum and related quantities in
experimental spectra. One is also curious whether the multi-source
thermal model describes appropriately the transverse flow.

In this paper, starting from the probability density function of
momenta in the relativistic form, we perform numerical
calculations in two steps to analyze the soft transverse momentum
($p_T$) spectra of the final state particles generated in high
energy gold--gold (Au--Au) and lead--lead (Pb--Pb) collisions
measured by the STAR~\cite{20,21,22,22a,22b} and ALICE
Collaborations~\cite{23,23a}, respectively. The thermodynamic
parameter $T$ (kinetic freeze-out temperature) and the
hydrodynamic parameter $\beta_T$ (transverse expansion velocity)
of the interacting system (emission source) are extracted. In the
first step, in the rest frame of emission source, using a
parameter $T$~\cite{24,25,26,27}, the momentum and its each
component and energy are sampled according to the assumption of
anisotropic emission and a given momentum distribution. In the
second step, a new transverse momentum distribution is obtained
according to the Lorentz transformation~\cite{28} at a given
$\beta_T$~\cite{28,29,30,31,32}.

The remainder of this paper is structured as follows. Section~2
introduces the related theoretical distribution and methods of
calculations. The comparison and discussion of the results are
presented in Section~3. Finally, in Section~4, we summarize and
list our main observations and conclusions.

\section{Formalism and method}

According to the multi-source thermal model~\cite{33,34}, many
emission sources are assumed to form in high energy collisions.
Each stationary emission source emits isotropically particles in
various directions. Furthermore, at the kinetic freeze-out, we
consider the relativistic ideal gas model~\cite{35,36,37} for the
escaped particles in the stationary source. The momentum ($p'$)
distribution of the emitted particles obeys the standard
distribution~\cite{37,38},
\begin{align}
f(p')=\frac{1}{N}\frac{dN}{dp'}=Cp'^2\left[
\exp\left(\frac{\sqrt{p'^2+m_{0}^{2}}-\mu}{T}\right)+
S\right]^{-1}
\end{align}
which is the probability density function. Here, $N$ is the number
of particles, $C=(1/m_{0}^{2}kT)[1/K_{2}(m_{0}/kT)]$ is the
normalization constant, $K_{2}(m_{0}/kT)$ is the modified
second-order Bessel function correction, $k=1$ is the Boltzmann
constant in the system of natural units, $T$ is the temperature of
the emission source, and $m_0$ and $\mu$ are the rest mass and
chemical potential of the particle, respectively. When $S=+1$, 0,
and $-1$, the function corresponds to the Fermi-Dirac
distribution, Boltzmann distribution, and Bose-Einstein
distribution, respectively~\cite{39}.

Generally, if $m_{0}\gg\mu$, the quantum effect plays a small
role, so we can ignore the influence of the chemical potential
when studying the momentum distribution in collisions at higher
energy. In case of $m_{0}\approx\mu$, the role of the chemical
potential increases significantly, and it is necessary for us to
distinguish the fermions and bosons. For the case of $m_0\ll\mu$,
the absolute value, $|\sqrt{p'^2+m_0^2}-\mu|$, should be used to
avoid $\sqrt{p'^2+m_0^2}-\mu<0$ in low-$p'$ region. In fact, $\mu$
is small enough in this work and the case of
$\sqrt{p'^2+m_0^2}-\mu<0$ does not exist. Although we may regard
$\mu$ as a free parameter, its value can be obtained in different
ways.

Empirically, for baryons, the chemical potential $\mu_B$ is given
by~\cite{40,41,42,43},
\begin{align}
\mu_{B}=\frac{1.303}{1+0.286\sqrt{s_{NN}}},
\end{align}
where $\sqrt{s_{NN}}$ is the collision energy (center-of-mass
energy) per nucleon pair, and both $\mu_{B}$ and $\sqrt{s_{NN}}$
are in GeV. Approximately, the chemical potential $\mu_p$ of the
proton is given by $\mu_B$, and the chemical potential $\mu_{\pi}$
($\mu_K$) of the pion (kaon) is taken from a method based on the
yield ratio of negative to positive particles~\cite{22b}.

For the convenience of subsequent calculations, we use the Monte
Carlo method~\cite{5,44,45} to obtain the momentum distribution
and discrete values of each component and energy. Let $R_{1,2,3}$
be random numbers that follow a uniform distribution in the range
of $[0,1]$. For a discrete $p'$, it satisfies the momentum
sampling:
\begin{align}
\int^{p'}_{0}f(p'')dp''<R_{1}< \int^{p'+\delta p'}_{0}f(p'')dp'',
\end{align}
where $\delta p'$ is a small shift from $p'$ and $f(p'')$ is just
Eq. (1). The azimuthal angle of the isotropic emission is obtained
by
\begin{align}
\varphi'=2 \pi R_{2},
\end{align}
and the emission angle is
\begin{align}
\vartheta'=2\arcsin\sqrt{R_{3}}.
\end{align}

Thus, in the rectangular coordinate system $O$-$xyz$, where $Oz$
axis is the beam direction and $xOz$ plane is the reaction plane,
one obtains the $x$-component of the momentum,
\begin{align}
p'_{x}=p'\sin\vartheta'\cos\varphi',
\end{align}
and the $y$-component of the momentum,
\begin{align}
p'_{y}=p'\sin\vartheta'\sin\varphi',
\end{align}
the transverse momentum,
\begin{align}
p'_{T}=p'\sin\vartheta',
\end{align}
the $z$-component of the momentum,
\begin{align}
p'_{z}=p'\cos\vartheta',
\end{align}
and the energy,
\begin{align}
E'=\sqrt{p'^{2}+m_{0}^{2}}.
\end{align}

Now we consider the Lorentz transformation from the stationary
source to the expanding one, which is caused by the interactions
among multiple sources. Let $\beta_{x,y,z}$ denote the components
of the expansion velocity. We have
\begin{align}
p_{x}=\frac{1}{\sqrt{1-\beta_{x}^{2}}}(p'_{x}+\beta_{x}E'),
\end{align}
\begin{align}
p_{y}=\frac{1}{\sqrt{1-\beta_{y}^{2}}}(p'_{y}+\beta_{y}E'),
\end{align}
and
\begin{align}
p_{z}=\frac{1}{\sqrt{1-\beta_{z}^{2}}}(p'_{z}+\beta_{z}E').
\end{align}
The transverse momentum is given by
\begin{align}
\begin{split}
p_{T}=&\sqrt{p_{x}^{2}+p_{y}^{2}}=\frac{1}{\sqrt{(1-\beta_{x}^{2})(1-\beta_{y}^{2})}}\\
&\times\sqrt{(1-\beta_{y}^{2})(p'_{x}+\beta_{x}E')^{2}+(1-\beta_{x}^{2})(p'_{y}+\beta_{y}E')^{2}}.
\end{split}
\end{align}
Here, $\sqrt{\beta_{x}^{2}+\beta_{y}^{2}}=\beta_{T}$. Due to the
small difference between $\beta_{x}$ and $\beta_{y}$, we ignore
this difference and take $\beta_{x}=\beta_{y}$ for simplicity when
describing the transverse momentum spectrum. It should be noted
that the difference cannot be ignored in analyzing anisotropic
flow.

It should be noted that Eq. (13) contains only the proper
expansion, but not the longitudinal expansion or motion of the
emission source. If the longitudinal effect is considered, we may
use a larger $\beta_z$ in Eq. (13). In this case, the ``expansion"
of emission source is anisotropic in the three dimensional
momentum space, though the expansion can be taken to be isotropic
in the transverse plane when the transverse anisotropic flow is
not the topic of the present work.

In the transverse plane, if the interactions among various sources
are considered, an anisotropic flow will be observed. Although one
may use $\beta_x \neq \beta_y$ to describe the transverse
anisotropic flow, which results in anisotropic $p_x$ and $p_y$
distributions, we may also have an alternative method. In our
previous work~\cite{45a}, to extract the property of transverse
anisotropic flow, one may appropriately revise Eqs. (11) and (12)
to be
\begin{align}
p_{x}=\frac{a_x}{\sqrt{1-\beta_{x}^{2}}}(p'_{x}+\beta_{x}E')+b_x
\end{align}
and
\begin{align}
p_{y}=\frac{a_y}{\sqrt{1-\beta_{y}^{2}}}(p'_{y}+\beta_{y}E')+b_y
\end{align}
respectively. Here, the parameters $a_x$ and $b_x$ ($a_y$ and
$b_y$) describe the source's expansion and displacement along $Ox$
($Oy$) axis respectively.

Generally, $a_{x,y}=1$ and $b_{x,y}=0$ imply an isotropic source.
For an anisotropic source, $a_{x,y}>1$ describes an expansion, and
$b_{x,y}>0$ ($b_{x,y}<0$) describes a displacement along the
positive (negative) direction~\cite{45a}. Although $a_{x,y}<1$ may
describe a contraction in mathematics, the physics condition gives
$a_{x,y}\geq 1$ in the expansion stage of the system. One may use
the elliptic flow
\begin{align}
v_2=\langle \cos(2\varphi) \rangle =\bigg\langle
\frac{p_x^2-p_y^2}{p_x^2+p_y^2} \bigg\rangle
\end{align}
to describe the strength of the transverse anisotropic flow, where
$\langle ...\rangle$ denotes an average over the considered data
sample. Comparing with $\beta_T$, the influence of $v_2$ on $p_T$
spectra is very small, which can be neglected in the following
analyses.

We would like to point out that, as a data-driven work, the
current analysis does not model any viscous hydrodynamics
approach. Instead, one may use the $p_T$ spectra to extract some
temperatures to describe the excitation degrees of the system at
the related stages, where the temperatures are free parameters. In
fact, the kinetic freeze-out temperature $T$ and transverse
expansion velocity $\beta_T$ of emission source used in this study
are set as free parameters. They describe the thermodynamic and
hydrodynamic characteristics of the interacting system at the
stage of kinetic freeze-out respectively.

\begin{figure*}[!htb]
\begin{center}
\vspace{-1cm}
\includegraphics[width=12.0cm]{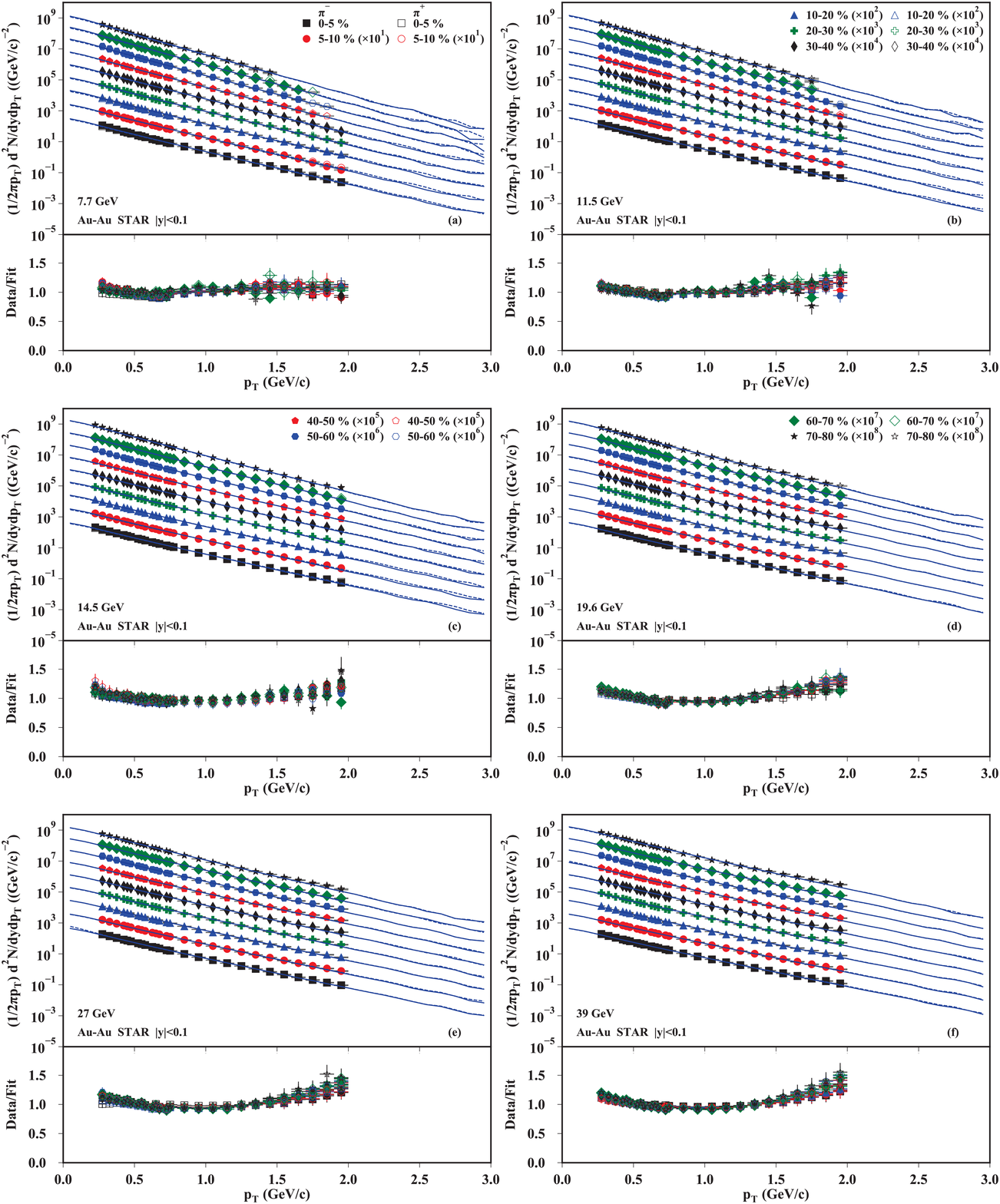}
\includegraphics[width=12.0cm]{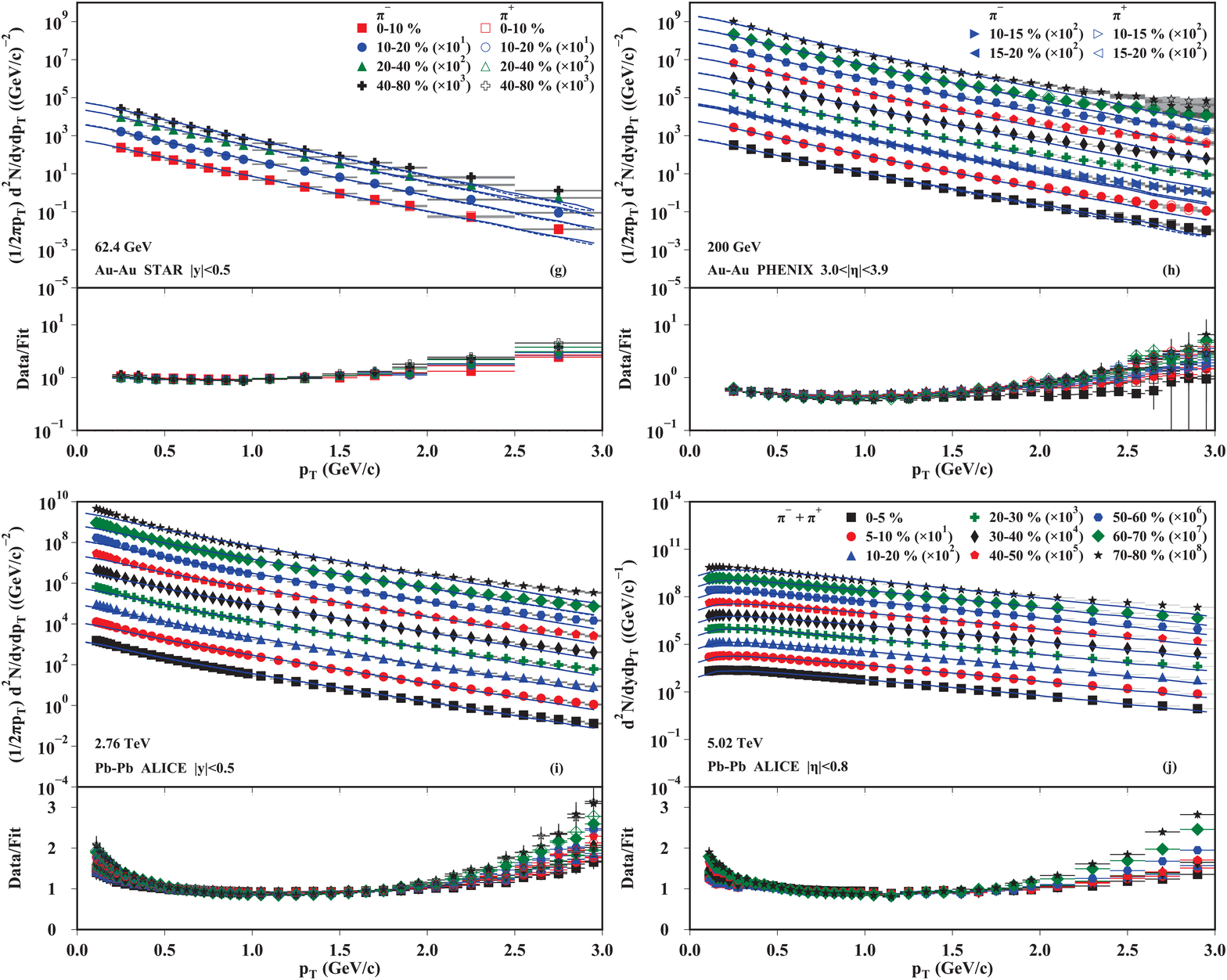}
\end{center}
\vskip-0.4cm\justifying\noindent {Figure~1. Transverse momentum
spectra of $\pi^{-}$ and $\pi^+$ produced in Au--Au collisions at
$\sqrt{s_{NN}}=7.7$~GeV (a), 11.5~GeV (b), 14.5~GeV (c), 19.6~GeV
(d), 27~GeV (e), 39~GeV (f), 62.4~GeV (g), and 200~GeV (h), as
well as in Pb--Pb collisions at 2.76~TeV (i) and 5.02 TeV (j) with
various centrality classes and given mid-rapidity. The symbols
represent the experimental data measured by the
STAR~\cite{20,21,22,22a,22b} and ALICE~\cite{23,23a}
Collaborations and re-scaled by different amounts marked in the
panels. The curves are our results fitted by the Bose-Einstein
distribution with embedded transverse expansion velocity.}
\end{figure*}

\begin{figure*}[!htb]
\begin{center}
\vspace{-1cm}
\includegraphics[width=12.0cm]{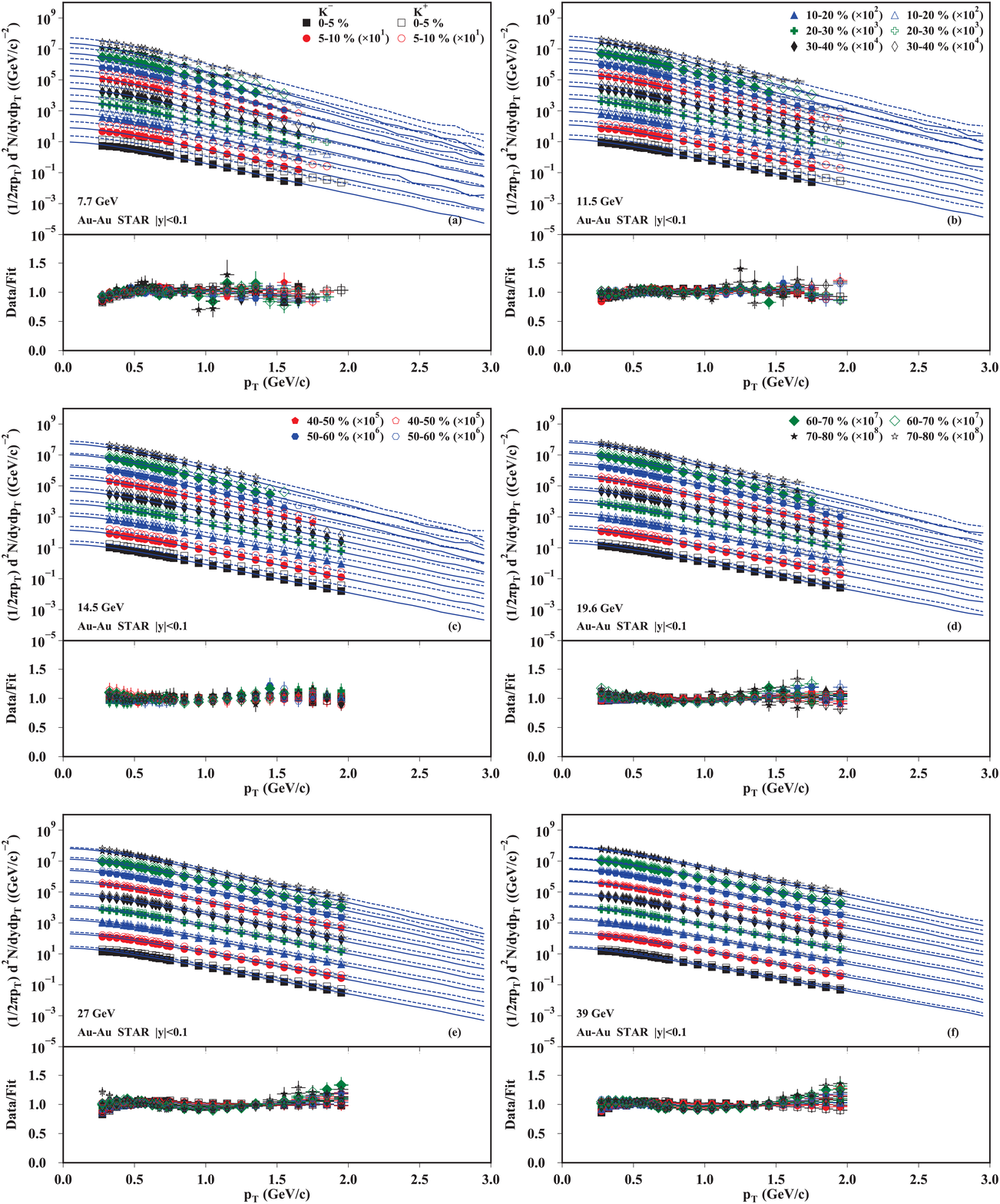}
\includegraphics[width=12.0cm]{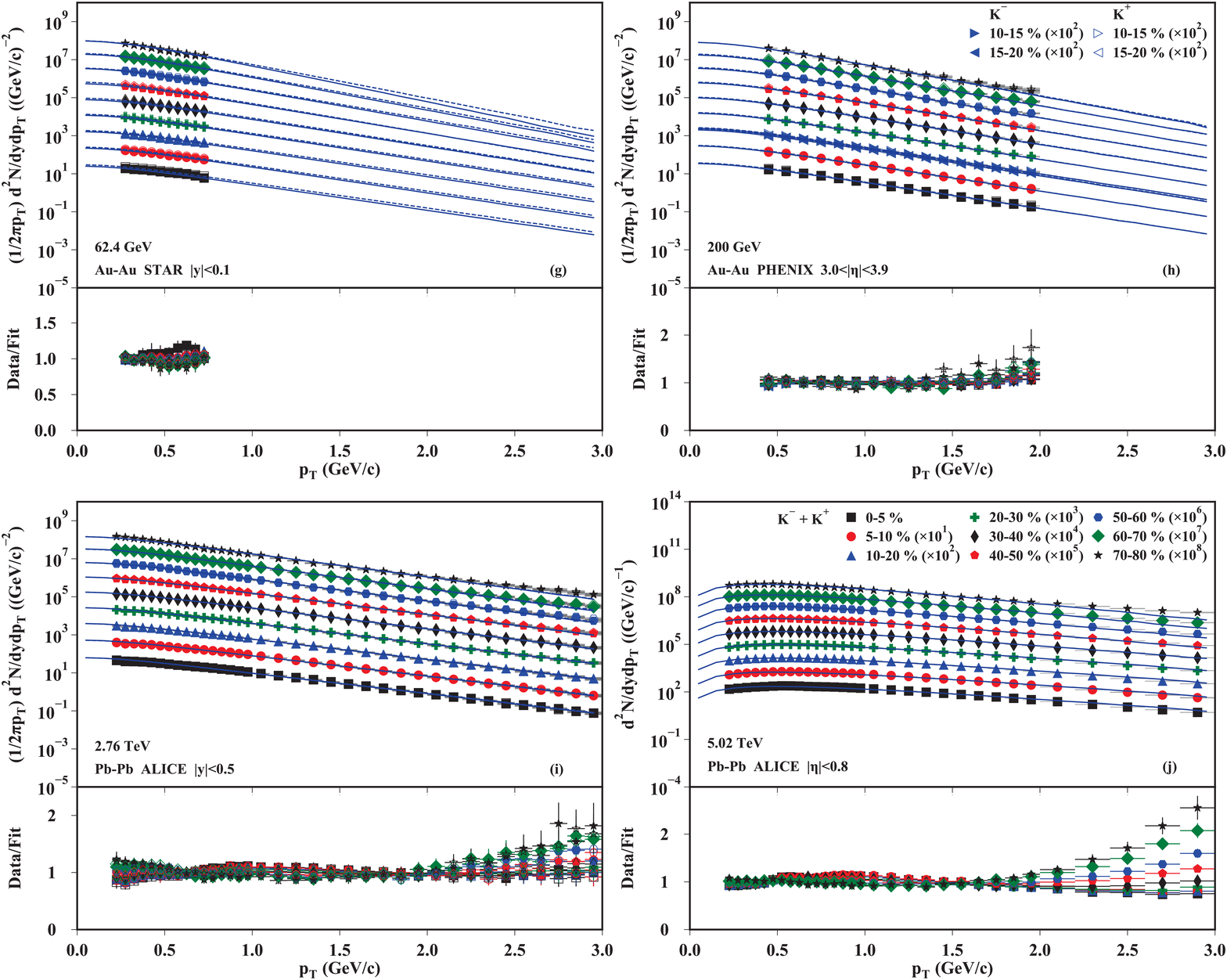}
\end{center}
\vskip-0.4cm\justifying\noindent {Figure~2. Same as Figure 1, but
showing the transverse momentum spectra of $K^{-}$ and $K^+$. The
curves are our results fitted by the Bose-Einstein distribution
with embedded transverse expansion velocity.}
\end{figure*}

\section{Results and discussion}

\subsection{Comparison with data}

To search for the energy of expected critical point of hadronic
matter transition to QGP, we may consider to study the
energy-dependent characteristics of thermodynamic and hydrodynamic
parameters, i.e. the excitation functions of these parameters.
This means that we have to collect and fit a large number of data
and display them in lots of figures. In fact, incomplete data
cannot give a whole excitation function, and the information
extracted from the incomplete excitation function may be
misleading. In the case of using lots of data and to make the
structure of this paper compact, we try to draw the same kind of
data in a figure.

Transverse momentum spectra of the final state particles generated
in Au--Au and Pb--Pb collisions over an energy range from 7.7 GeV
to 5.02 TeV, measured by the STAR~\cite{20,21,22,22a,22b} and
ALICE~\cite{23,23a} Collaborations, and then performed a fitting
analysis on them are displayed in Figures 1--3. In order to
compare the properties of the $p_{T}$ spectra at all energies more
comprehensively, we limit the range of $p_{T}$ to be equal to or
less than 3~GeV$/c$, which falls into the soft $p_T$ region.
Concretely, in the fit, the spectra in the range of $0.2<p_T<2.5$
GeV/$c$ is used to exclude the contributions of resonance decay
and hard scattering process~\cite{22,22a,22b} as much as possible.
Different symbols represent the experimental data in different
centrality classes, and the curves are our results calculated by
the Monte Carlo method. In addition, for showing $p_T$ spectra
more clearly in different centrality classes, we indented the data
to the power of 10 (such as the numbers in parentheses in the
legend). The length of the gray rectangle in the figure represents
the error of $p_{T}$, and the width represents the error of
$p_{T}$ spectrum $[(1/2\pi p_{T})d^{2}N/dydp_{T}]$, which is the
quadratic sum of the statistical and systematic uncertainties.

Figure~1 shows the result of fitting analysis of $p_{T}$ spectra
of negatively and positively charged pions ($\pi^{-}$ and $\pi^+$)
from central (0--5\% or 0--10\%) to peripheral (70--80\% or
40--80\%) collisions using the Bose-Einstein distribution embedded
an expansion velocity. Panels (a)--(j) represent the $p_{T}$
spectra at different collision energies shown in the figure. The
closed (open) symbols and solid (dashed) curves correspond to
$\pi^-$ ($\pi^+$), and there is no distinction between $\pi^-$ and
$\pi^+$ in panel (j). The experimental data presented in panels
(a)--(h) are quoted from the STAR Collaboration for Au--Au
collisions, which shows the mid-rapidity range of
$|y|<0.1$~\cite{20,21,22,22a,22b}; and the experimental data
presented in panels (i) and (j) are collected from ALICE
Collaboration for Pb--Pb collisions, whose mid-(pseudo)rapidity
range is $|y|<0.5$~\cite{23} and $|\eta|<0.8$~\cite{23a}. The
values of the free parameters $T$ and $\beta_{T}$ extracted from
the fits will be analyzed in next subsection. Following each
panel, the ratios of Data/Fit are given to show the quality of the
fit.

Similar to Figure~1, Figures~2 and 3 show the $p_T$ spectra of
negative and positive kaons ($K^{-}$, $K^+$, or $K^{-}+K^+$) and
anti-protons and protons ($\bar{p}$, $p$, or $\bar{p}+p$)
respectively~\cite{20,21,22,22a,22b,23,23a}, where the spectra in
Figure 2 are fitted with the Bose-Einstein distribution with
embedded transverse expansion velocity and the spectra in Figure 3
are fitted with the Fermi-Dirac distribution with embedded
transverse expansion velocity. The related parameters will be
analyzed in next subsection.

From Figures 1--3 one can see that the Bose-Einstein/Fermi-Dirac
distribution with embedded transverse expansion velocity can
describe the soft $p_T$ spectra of light charged particles
produced in high-energy Au--Au and Pb--Pb collisions at the
Relativistic Heavy Ion Collider and the Large Hadron Collider. In
the description, there are only two free parameters: the kinetic
freeze-out temperature $T$ and the transverse expansion velocity
$\beta_T$. Compared with the general treatments in the community,
the present work has used a more clear picture and a simpler
distribution. In fact, the Bose-Einstein/Fermi-Dirac distribution
is the most basic one in the ideal gas model.

From the ratios of Data/Fit, one can see that in very low $p_T$
region and around $p_T\approx3$ GeV/$c$, the fitting results
underestimate the data in some cases. This means that we need a
very soft component for the contribution of resonance decay.
Meanwhile, we need a hard component for the contribution of hard
scattering process. Naturally, a multi-component distribution can
be applied for the whole $p_T$ region. Generally, for the
extraction of thermodynamic and hydrodynamic characteristics of
interacting system, the contributions of resonance decay and hard
process to the $p_T$ spectra should be removed because the two
contributions come from non-thermal production and undergo
non-expansion process. Although other methods can describe the
$p_T$ spectra, the present work provides an alternative method in
the description. It does have clear picture and methodological
significance.

\subsection{Tendencies of parameters}

To show the excitation functions of related parameters and extract
the variation tendencies of the functions, we analyze the
energy-dependent relations of related parameters for different
centralities and particles in this subsection. The excitation
functions are related to the search for the energy of expected
critical point and the change of interaction mechanism, which
shows remarkable scientific significance. In addition, we have
used a method closer to the relativistic ideal gas
model~\cite{35,36,37} than the blast-wave model~\cite{21,23,29} in
fitting the transverse momentum spectra. It is also interesting to
compare the results of two fittings.

Figure~4 shows the dependence of $T$ on the collision energy
$\sqrt{s_{NN}}$, centrality class (percentage), and particle mass.
From panels (a) to (i), the centrality classes are 0--5\%,
5--10\%, ..., 70--80\%, respectively. The centrality classes at
62.4 and 200 GeV are different in some cases which shows in the
panels. The closed (open) squares, circles, and triangles
represent the results for $\pi^-$ ($\pi^+$), $K^-$ ($K^+$), and
$\bar p$ ($p$), respectively. The weighted averages over the
yields of different particles are shown by the crosses. One can
see that $T$ increases generally with some fluctuations as the
increase of collision energy. From central to peripheral
collisions, $T$ has a slight decrease. In central collisions, $T$
increases with the increase of particle mass, and in peripheral
collisions the situation seems to be different or ambiguous. The
dependence of $T$ on isospin is not significant.

Figure 5 is similar to Figure 4, but it shows the dependence of
$\beta_T$ on the collision energy, centrality class, and particle
mass. One can see that $\beta_T$ increases generally with some
fluctuations as the increase of collision energy. From central to
peripheral collisions, $\beta_T$ has a slight decrease. In central
collisions, $\beta_T$ does not change significantly with the
increase of particle mass, and in peripheral collisions, $\beta_T$
decreases with the increase of particle mass. The dependence of
$\beta_T$ on isospin is also not significant. Compared with $T$,
$\beta_T$ shows larger fluctuations in the spectra, which is
reflected by its errors.

It should be noted that the dependence of $T$ on centrality is an
open question at present, though $\beta_T$ increases with the
increase of centrality. Based on the blast-wave model~\cite{29},
the STAR~\cite{20,21} and ALICE~\cite{23} Collaborations shows
that $T$ decreases slightly with the increase of centrality.
However, our results show that $T$ increases slightly with the
increase of centrality. The difference is caused by different
methods which also lead to different values in parameters. In the
blast-wave model~\cite{29}, a self-similar and variant flow
profile function is used. In the present work, we have used an
invariant flow velocity embedded in the relativistic ideal gas
model~\cite{35,36,37} for the little bang process of relativistic
collisions, which is meaningful in methodology.

In central collisions, although a lower temperature can be
explained by a longer lifetime of the system, a higher temperature
can be explained by a higher excitation degree. The result of the
blast-wave model fitted by the STAR Collaboration~\cite{21} shows
that $T$ is almost the same in the energy range from 7.7 to 39 GeV
and it is lower at 62.4 and 200 GeV. Our result shows that $T$
increases generally from 7.7 GeV to 5.02 TeV. Although the results
of both the fittings can be explained by us, it is hard to
determine which one is correct. In our opinion, more model work is
needed in the future for the complex process of relativistic
collisions.

\begin{figure*}[!htb]
\begin{center}
\vspace{-1cm}
\includegraphics[width=12.0cm]{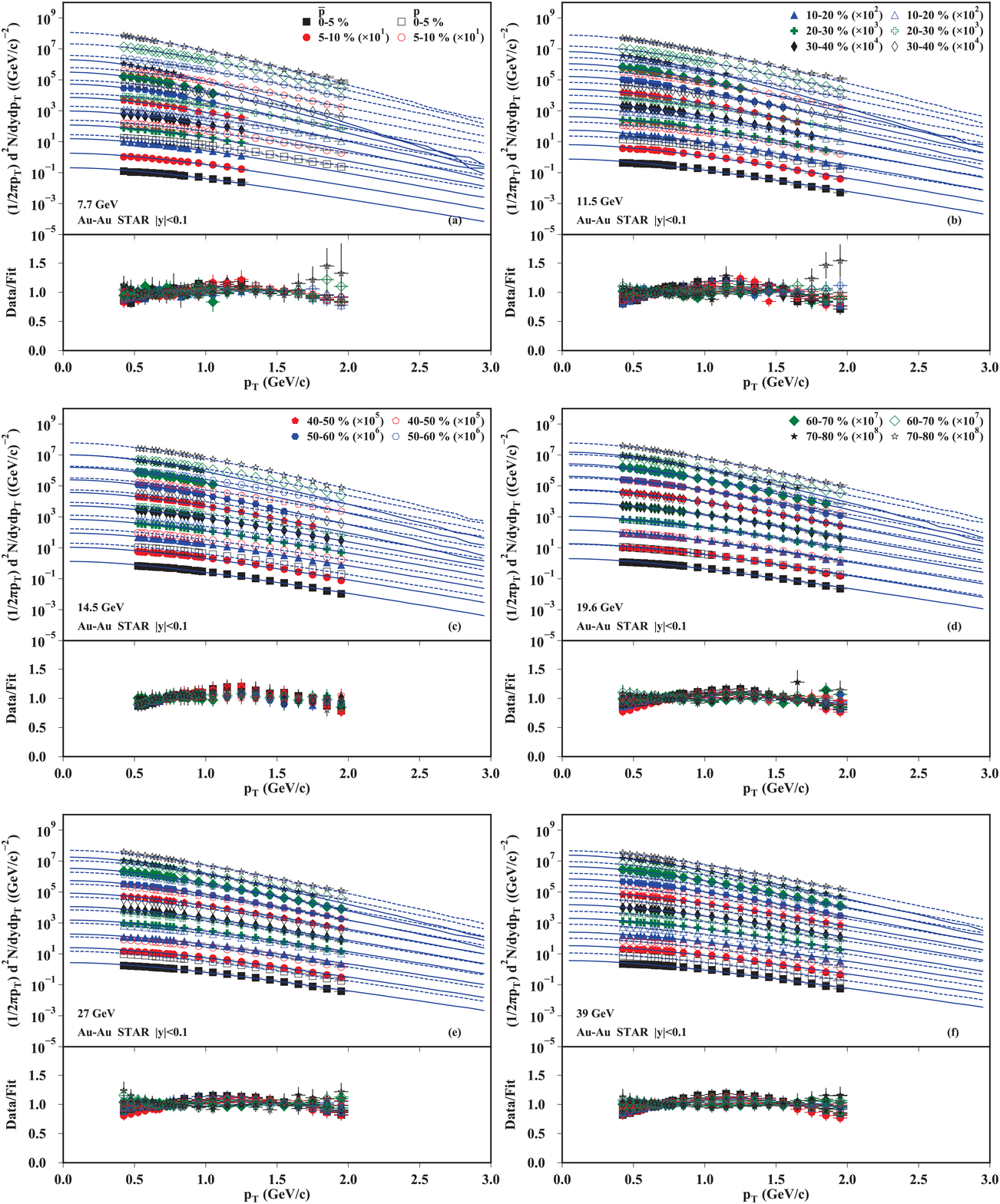}
\includegraphics[width=12.0cm]{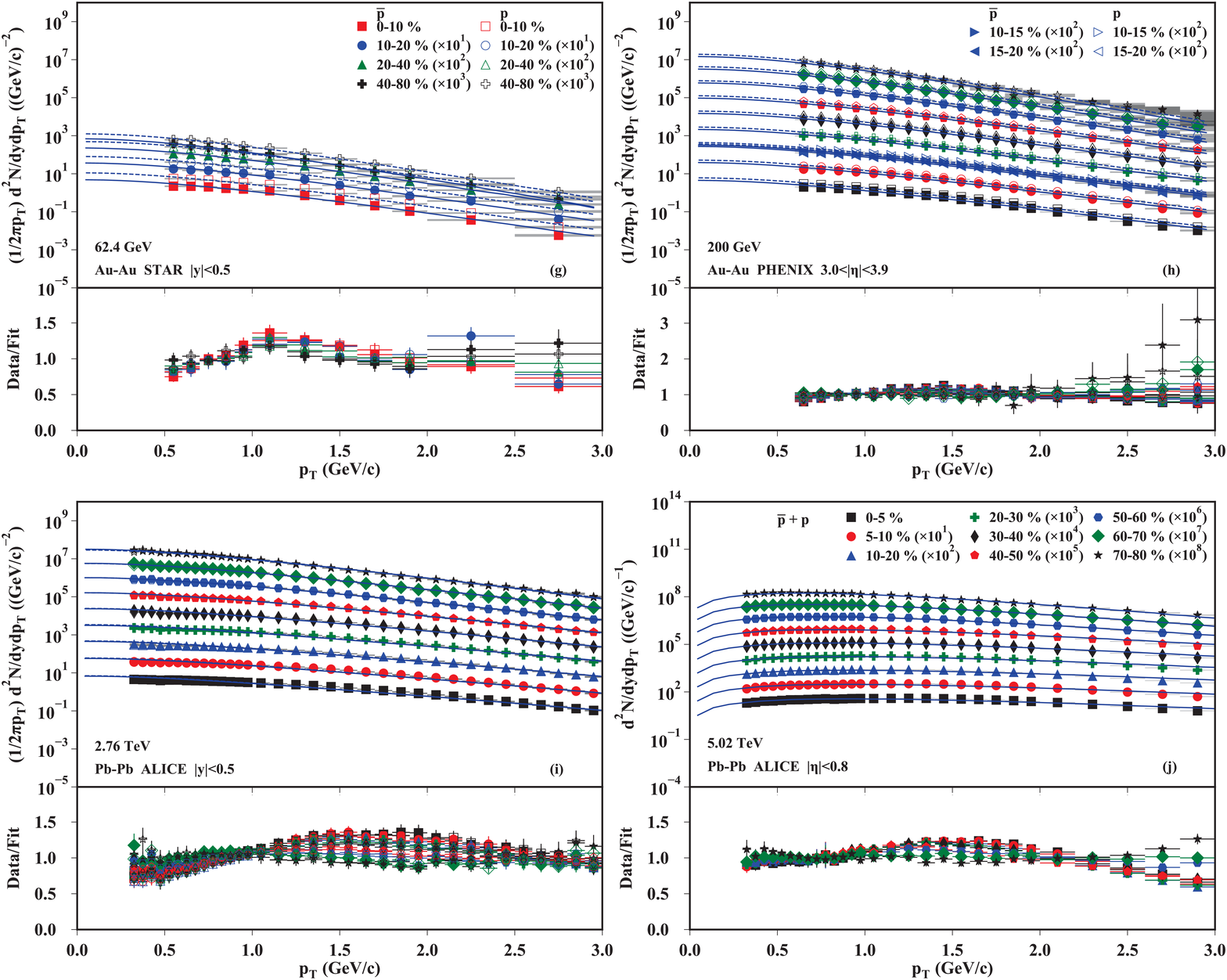}
\end{center}
\vskip-0.4cm\justifying\noindent {Figure~3. Same as Figure 1, but
showing the transverse momentum spectra of $\bar{p}$ and $p$. The
curves are our results fitted by the Fermi-Dirac distribution with
embedded transverse expansion velocity.}
\end{figure*}

\begin{figure*}[!htb]
\begin{center}
\includegraphics[width=15.0cm]{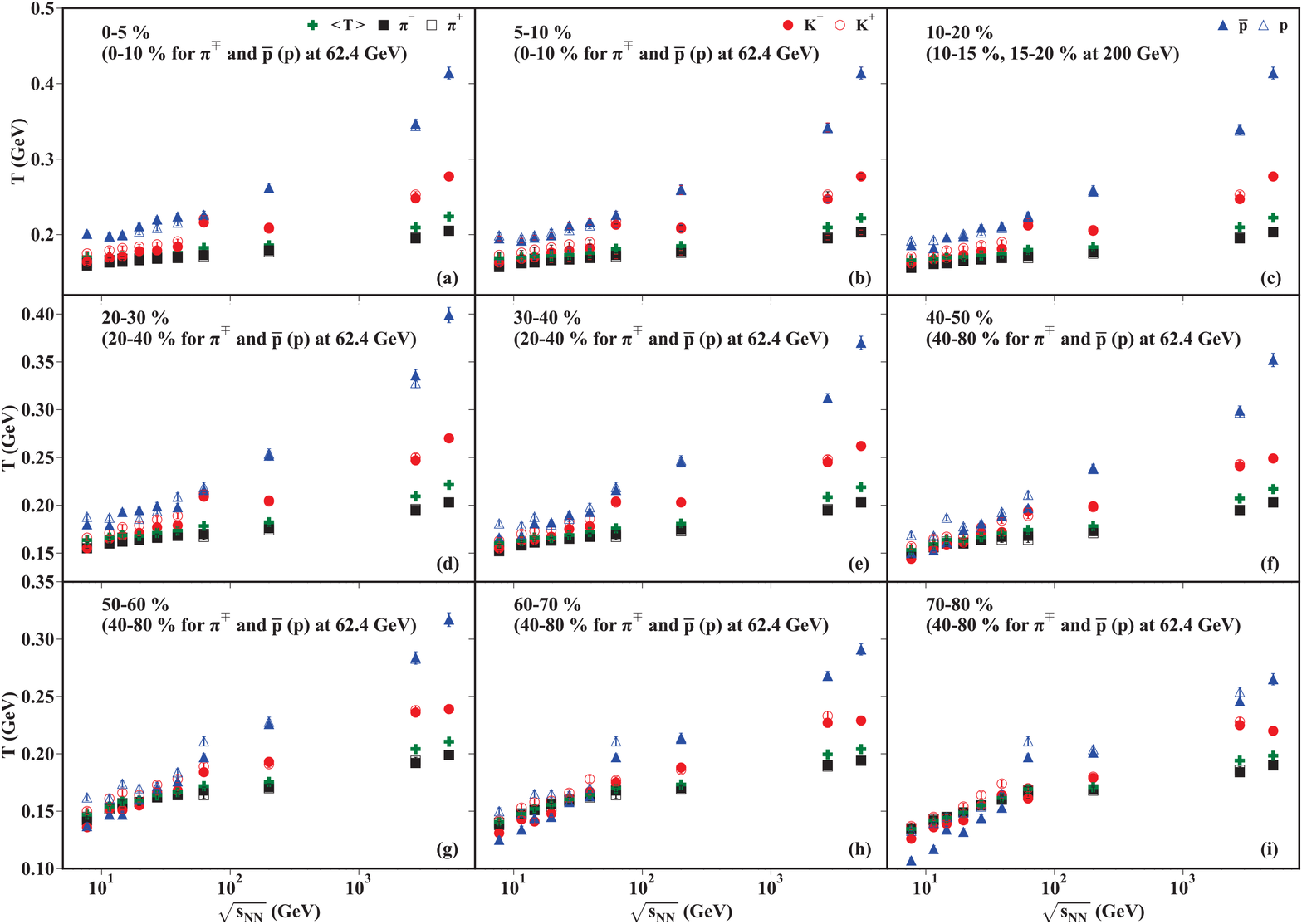}
\end{center}
\vskip-0.4cm \justifying\noindent {Figure~4. Dependence of the
kinetic freeze-out temperature $T$ on the collision energy
$\sqrt{s_{NN}}$, centrality, and particle mass or type. From
panels (a) to (i), the centrality classes are mainly 0--5\%,
5--10\%, ..., and 70--80\%, respectively. The closed (open)
squares, circles, and triangles represent the results for $\pi^-$
($\pi^+$), $K^-$ ($K^+$), and $\bar p$ ($p$), respectively, where
the closed symbols at 5.02 TeV are not undistinguished the
charges. The crosses represent the average $T$ ($\langle
T\rangle$) weighted over the yields of different particles. All
symbols represent the results fitted from Figures 1--3.}
\begin{center}
\includegraphics[width=15.0cm]{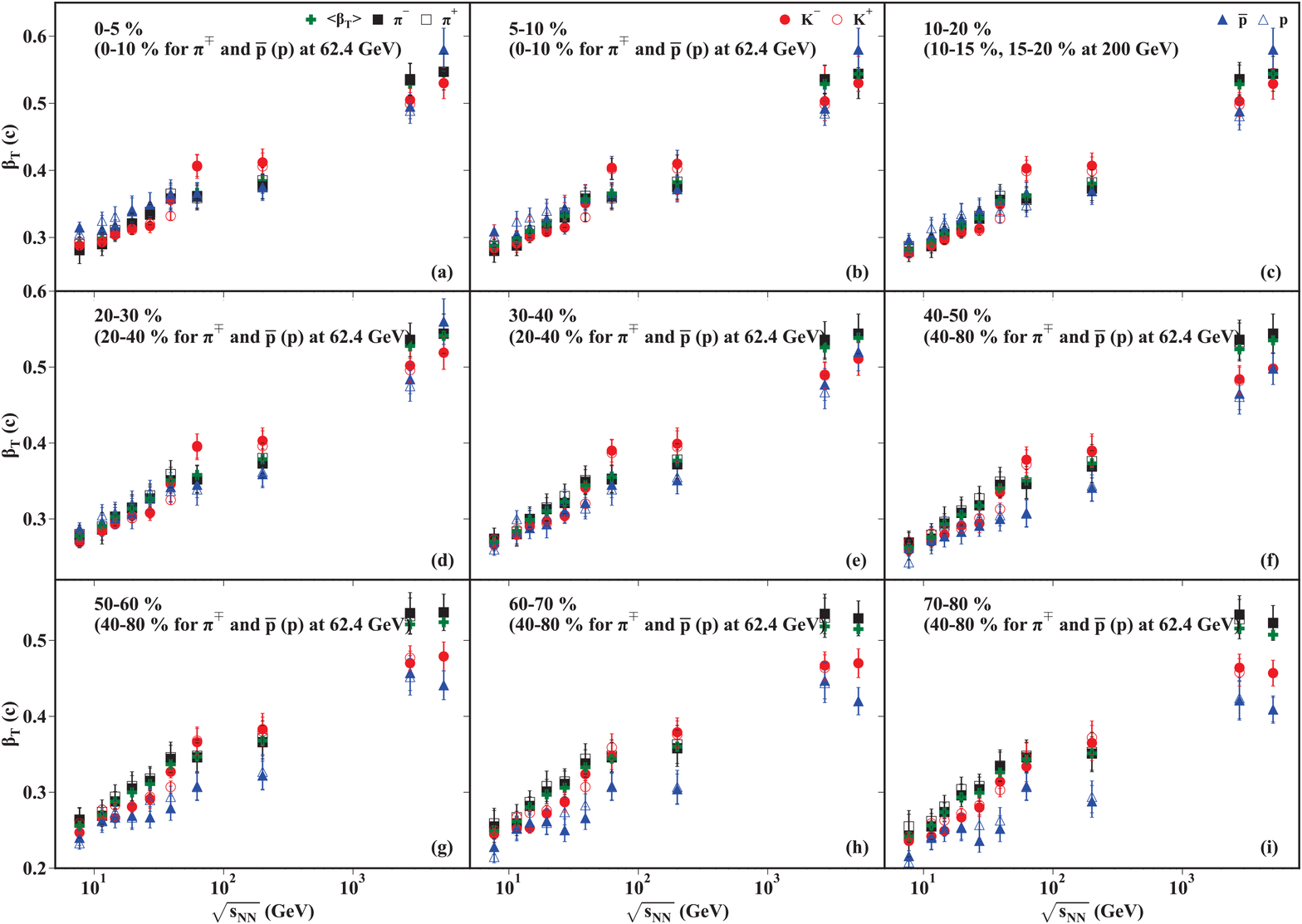}
\end{center}
\vskip-0.4cm \justifying\noindent {Figure~5. Same as Figure 4, but
showing the dependence of the transverse expansion velocity
$\beta_T$ on the collision energy $\sqrt{s_{NN}}$, centrality, and
particle mass. The average $\beta_T$ ($\langle \beta_T\rangle$) is
the weighted average over the yields of different particles.}
\end{figure*}

\subsection{Further discussions}

Generally, the interacting system of relativistic heavy ion
collisions~\cite{15,16} will experience two stages of freeze-out,
one is chemical freeze-out and the other is kinetic freeze-out.
The two kinds of freeze-outs possibly occur simultaneously or
non-simultaneously. It depends on the sizes of chemical and
kinetic freeze-out temperatures. In fact, the system is suddenly
frozen~\cite{46,47} and the short-lived resonances
decay~\cite{48}, changing the kinetic spectra of stable particles.
The resonances of the system, generated during chemical
freeze-out, decay rapidly, but the system continues to evolve with
elastic collisions between hadrons, and the system will stay in
the local thermal equilibrium before kinetic freeze-out~\cite{49}.
The particles' transverse momentum spectra carry information about
it.

If the chemical and kinetic freeze-out temperatures are nearly the
same, one may consider that the two freeze-outs occur almost
simultaneously. If the two freeze-outs occur at different time
moments, the chemical freeze-out happens generally earlier than
the kinetic freeze-out, and the chemical freeze-out temperature is
larger than the kinetic one. The temperature $T$ extracted from
the present work is only the kinetic freeze-out temperature. The
chemical freeze-out temperature is not available here, and we
cannot give a comparison for the two temperatures corresponding to
the two freeze-outs.

Studying the traverse momentum (mass) spectra of the final state
particles, produced in relativistic heavy ion collisions, is an
effective and fast means to obtain thermodynamic parameters of the
system. The emission source determines the transverse momentum
spectra of different kinds of particles. When the interacting
system is in the stage of kinetic freeze-out, the emitted
particles not only contain thermal motion, but also are affected
by the expansion of the system or the flow velocity of the
particles. Thermal motion reflects the transverse excitation
degree of the system, which can be reflected by the kinetic
freeze-out temperature $T$. The expansion or flow effect embodies
the hydrodynamic feature of the system, which can be represented
by the transverse expansion or flow velocity $\beta_{T}$.

In the community, the transverse momentum spectra have been
studied extensively for different final state particles. The
functions usually used in fitting the transverse momentum spectra
include the Tsallis distribution~\cite{50,51,52}, the Erlang
distribution~\cite{53,54,55}, the Hagedorn function~\cite{56,57},
etc. In the present work, we have adopted the most basic function,
the Bose-Einstein/Fermi-Dirac distribution, in the relativistic
ideal gas model and introduced the transverse expansion velocity
of the system to analyze and fit the transverse momentum spectra
of the charged particles, $\pi^{-}$, $\pi^{+}$, $K^{-}$, $K^{+}$,
$\bar{p}$, and $p$ produced in Au--Au and Pb--Pb collisions at
high energies, and obtain the thermodynamic parameter $T$ and the
hydrodynamic parameter $\beta_T$.

Our work shows that the excitation functions of $T$ and $\beta_T$
increase with the increase of collision energy in the concerned
range from 7.7 GeV to 5.02 TeV. This implies that the excitation
and expansion degrees of the interacting system are higher at
higher energy. The interaction mechanism in the concerned energy
range should be the same, which involves the formation of hot
dense matter. There is no obvious evidence for the energy of
critical point being observed from this work. The energy of
expected critical point of hadronic matter transition to QGP is
possibly in the lower energy range. Because the most basic
function, the Bose-Einstein/Fermi-Dirac distribution, in the
relativistic ideal gas model is used, and the transverse expansion
velocity of the system is introduced, we consider that the present
work has a more solid foundation than the other distributions or
functions used in the community. Moreover, there are only two free
parameters in the description of transverse momentum spectra,
which is also an advantage of the present work.

It should be noted that the present work shows different
tendencies of parameters on collision energy and centrality from
the blast-wave model~\cite{29} used by the STAR and ALICE
Collaborations~\cite{20,21,23}. The reason is that different
pictures and $p_T$ ranges are used in the two methods. Except for
the invariant $\beta_T$ used in the present work and variant
$\beta_T$ used in the STAR and ALICE Collaborations, the present
work uses the data in a wider $p_T$ range available in experiments
and the STAR and ALICE Collaborations used the data in narrow
intermediate-$p_T$ range. For example, the STAR Collaboration
pointed out that ``the fit ranges used for pions, kaons, and
protons are 0.5--1.35 GeV/$c$, 0.3--1.35 GeV/$c$, and 0.5--1.25
GeV/$c$, respectively", for central Au--Au collisions at 14.5
GeV~\cite{21}.

Indeed, as mentioned by the STAR and ALICE
Collaborations~\cite{21,23}, ``the blast-wave model fits are very
sensitive to the $p_T$ range used". However, this is not the case
of the present work. Although the resonance decays contribute in
low-$p_T$ range and the hard scattering process contributes mainly
in high-$p_T$ range, it is hard to separate various ranges
completely. In fact, the soft thermal process has also large
probability to contribute in low-$p_T$ range and small probability
to contribute in high-$p_T$ range. In a wider $p_T$ range, the
proportion of low-$p_T$ is very large, though that of high-$p_T$
range is very small.

In our opinion, we should use the $p_T$ range as wide as possible
in the extraction of kinetic freeze-out parameters from the
spectra of soft process. This means that we have to give a
consideration to exclude the contributions of resonance decay and
hard process to the $p_T$ spectra. However, the boundary in the
spectra of different processes is not completely separate. After
weighing the pros and cons~\cite{22,22a,22b}, we take
$p_T=0.2$--2.5 GeV/$c$ in this work.

Before summary and conclusions, we would like to point out that
the equation of state effects (for the interacting medium) on the
$p_T$ spectra are not considered into the analysis. The reason is
that the assumption of isotropic stationary emission source is
applicable, which implies a small influence of the mentioned
effects. However, if one studies the anisotropic elliptic flow,
the interactions among various sources should be considered. This
means that the density of the interacting medium is varying, and
then the pressure, temperature, viscosity, and other quantities
are changeable in different local regions. These changes are
partly reflected by anisotropic elliptic flow which can be
described by $a_{x,y}$ and $b_{x,y}$, or different $\beta_x$ and
$\beta_y$, in the multi-source thermal model appropriately.

\section{Summary and Conclusions}

Based on the framework of a multi-source thermal model, we have
analyzed the soft transverse momentum spectra of $\pi^{-}$,
$\pi^{+}$, $K^{-}$, $K^{+}$, $\bar{p}$, and $p$ produced in Au--Au
collisions at $\sqrt{s_{NN}}=7.7$, 11.5, 14.5, 19.6, 27, 39, 62.4,
and 200~GeV, measured by the STAR Collaboration, and in Pb--Pb
collisions at $\sqrt{s_{NN}}=2.76$ and 5.02~TeV, measured by the
ALICE Collaboration. In the rest framework of emission source, the
probability density function of meson momenta satisfies the
Bose-Einstein distribution, and that of baryon momenta satisfies
the Fermi-Dirac distribution.

Considering the interactions among multiple sources, the emission
source has an expansion velocity or the particles have a flow
velocity. To simulate the transverse momentum spectra, the kinetic
freeze-out temperature and transverse expansion velocity of
emission source are introduced. The numerical results, calculated
by the Monte Carlo method, are in good agreement with the
experimental data of the STAR and ALICE Collaborations. The
excitation function, and the centrality and particle mass
dependences of kinetic freeze-out temperature and transverse
expansion velocity are obtained from the analyses. Being the most
basic function, the Bose-Einstein/Fermi-Dirac distribution with
the introduced transverse expansion velocity implies that the
present work has a solid foundation.

This work does not support the current view that the kinetic
freeze-out temperature decreases with the increase of collision
energy in the considered energy range. In fact, our work shows
that the kinetic freeze-out temperature and the transverse
expansion velocity increase generally with the increase of
collision energy. The excitation functions of the two parameters
do not show a particular structure. This implies that the
interaction mechanism in the concerned energy range is the same,
though the degrees of excitation and expansion of the system are
higher at higher energy. The same mechanism involves the formation
of hot dense matter. If existing, the energy of expected critical
point of hadronic matter transition to quark-gluon plasma is below
the present energy range.

This work also does not support the current view that the kinetic
freeze-out temperature in peripheral collisions is higher than
that in central collisions. On the contrary, our work shows that
with the increase of centrality from peripheral to central
collisions, both the kinetic freeze-out temperature and the
transverse expansion velocity have a slight increase. This implies
that the degrees of excitation and expansion of the system in
central collisions are higher. The reason is that more energy are
deposited in central collisions due to the involvement of multiple
nucleons.

In addition to energy and centrality dependences, with the
increase of particle mass, the kinetic freeze-out temperature
increases and the transverse expansion velocity decreases in some
cases. This reflects the characteristics of the evolution of
hydrodynamic system, in which massive particles are leaved in the
early stage at high temperature.

When looking forward into the future, in order to search for the
critical point of deconfinement phase transition from hadronic
matter to QGP, it is very necessary to study the process of high
energy collisions in lower energy region. The several ongoing
heavy ion experiments at a few GeV performed around the world will
answer this issue. We look forward to the new results.
\\
\\
{\bf Acknowledgments}

The work of X.-H.Z. was supported by the Innovative Foundation for
Graduate Education in Shanxi University. The work of Shanxi Group
was supported by the National Natural Science Foundation of China
under Grant No. 12147215, the Shanxi Provincial Natural Science
Foundation under Grant No. 202103021224036, and the Fund for
Shanxi ``1331 Project" Key Subjects Construction. The work of
K.K.O. was supported by the Agency of Innovative Development under
the Ministry of Higher Education, Science and Innovations of the
Republic of Uzbekistan within the fundamental project No.
F3-20200929146 on analysis of open data on heavy-ion collisions at
RHIC and LHC.
\\
\\

\end{document}